%% file: main.tex
\documentclass[conference]{IEEEtran}
\IEEEoverridecommandlockouts

\usepackage{cite}
\usepackage{amsmath,amssymb,amsfonts}
\usepackage{algorithmic}
\usepackage{graphicx}
\usepackage{textcomp}
\usepackage{xcolor}
\usepackage[nolist]{acronym}
\newcommand{\fref}[1]{Fig.~\ref{#1}}

\newcommand{\tref}[1]{Table~\ref{#1}}

\usepackage[caption=false]{subfig}
\usepackage{tikz}
\tikzset{
  font={\fontsize{9pt}{12}\selectfont}}
\usetikzlibrary{plotmarks}
\usetikzlibrary{intersections,backgrounds, patterns}
\usetikzlibrary{patterns,shapes.arrows}
\usepackage{balance}
\usepackage{pgfplots}
\usepackage{pgfplotstable}
\pgfplotsset{compat=newest} 
\usepgfplotslibrary{groupplots,dateplot}
\pgfplotsset{every axis/.append style={
	scaled x ticks = false,
	label style={font=\footnotesize},
	tick label style={font=\footnotesize},
	tick scale binop=\times}
}
\pgfkeys{/pgf/number format/.cd,
1000 sep={},
}

\usepackage{booktabs}
\usepackage{amsmath}
\usepackage{multirow}
\definecolor{hgreen}{rgb}{0, 0.5, 0}

\def\BibTeX{{\rm B\kern-.05em{\sc i\kern-.025em b}\kern-.08em
    T\kern-.1667em\lower.7ex\hbox{E}\kern-.125emX}}
\begin{document}

\title{
Disaggregated multi-domain interference classification for O-RAN \\
\thanks{This research was supported by the 6G-Bricks project (Grant Agreement No. 101096954) and the Sunrise-6G project (Grant Agreement No. 101139257), both co-funded by the European Union under its Horizon Europe Research and Innovation Program.}
}

\author{\IEEEauthorblockN{Dieter Verbruggen, Hazem Sallouha, and Sofie Pollin}
\IEEEauthorblockA{\textit{Department of Electrical Engineering (ESAT) - WaveCoRE}\\
KU Leuven, 3000 Leuven, Belgium\\
E-mail:\{dieter.verbruggen, hazem.sallouha, sofie.pollin\}@kuleuven.be}}

\maketitle

\input{0.abstract}

\begin{IEEEkeywords}
O-RAN, Interference classification, Distributed Processing, Network Architecture, Deep-Learning
\end{IEEEkeywords}

\input{1.Introduction}
\input{2.system_architecture}
\input{3.Results}
\input{4.Conclusion}
\bibliographystyle{ieeetr}
\bibliography{bibliography}

\end{document}

%% file: 0.abstract.tex
\begin{abstract}
Spectrum sharing and dynamic spectrum reuse are becoming increasingly critical in modern wireless networks to address spectrum scarcity. However, these techniques inevitably increase Cross-Technology Interference (CTI). In this context, the Open Radio Access Network (O-RAN), as a modern and disaggregated network architecture, necessitates accurate, low-latency, and computationally efficient CTI classification and mitigation to support real-time control and maintain Quality of Service (QoS). Unfortunately, existing solutions predominantly rely on high-complexity, monolithic deep learning-based solutions that, while achieving high classification accuracy, incur significant latency and computational overhead
This paper exploits the O-RAN functional split to leverage multi-domain raw signal representations (time, frequency, and Channel State Information (CSI)) directly from the same data stream. Each domain is processed locally, naturally interleaving CTI within the distributed, disaggregated O-RAN architecture. This distributed strategy enables a cost-aware, multi-domain fusion architecture that balances classification accuracy with computational overhead and latency. 
Our proposed multi-domain distributed architecture achieves a 400 $\mu s$ inference latency on standard CPUs. Compared to a state-of-the-art monolithic frequency-domain classifier, this represents an average 9x reduction in latency and an 11-fold decrease in computational cost, while sacrificing only 4\% in classification performance and maintaining $>$ 90\% accuracy in high-interference conditions


\end{abstract}

%% file: 1.Introduction.tex
\section{Introduction}
The increasing demand for wireless connectivity is placing pressure on the radio spectrum. To improve spectrum utilization, modern wireless systems increasingly rely on spectrum sharing, dynamic reuse, and heterogeneous deployments \cite{spectrum}. While these techniques improve spectral efficiency, they also increase the occurrence of \ac{cti}, in which signals from different wireless technologies coexist in overlapping spectral resources \cite{cti,inference_mit}. Accurate and timely identification of such interference is therefore essential for maintaining Quality of Service (QoS) in next-generation architectures, such as those built on the emerging \ac{oran} standard.

The architecture of \ac{oran} facilitates intelligent interference management by providing direct access to intermediate \ac{phy} signal representations \cite{oran_understanding}. Under the functional Split 7.2x \cite{oran_wg4_fronthaul}, the physical layer is divided between the \ac{ru} (Low-PHY) and the \ac{du} (High-PHY). This division yields data across three distinct domains: time-domain samples at the \ac{ru}, alongside frequency-domain symbols and \ac{csi} estimates computed at the \ac{du}. Parallel to this data pipeline operates a multi-tier control plane governed by distinct latency loops. This plane hosts control applications spanning multiple timescales, such as rApps for non-real-time policy management, xApps for near-real-time control, and \acp{dApp} for real-time tasks such as spectrum sensing,  scheduler reconfiguration, and interference management \cite{dapps}. Deployed as a \ac{dApp}, \ac{cti} classification must operate within a shared physical compute infrastructure alongside other performance-critical applications. This shared environment, coupled with the need to react to spectral changes within the strict 1 ms 5G subframe \cite{dahlman20205g}, makes the deployment of a lightweight \ac{cti} model classification essential. 

Conventionally, a \ac{cti} \ac{dApp} executes on the \ac{du}, relying on frequency-domain symbols or post-equalized data transformed by the high-PHY pipeline \cite{inferoran}. However, disaggregating the classification model across the \ac{oran} Split 7.2x pipeline offers a significant advantage. By performing early time-domain analysis directly at the \ac{ru} and fusing the resulting latent features with frequency and \ac{csi} representations at the \ac{du}, the model exploits complementary data without imposing additional signal-processing overhead. This structural shift naturally facilitates domain fusion, which enhances model robustness across diverse wireless channels and \ac{snr} \cite{costmulti2, zhengsnr}. Applying this multi-domain methodology specifically to \ac{cti} \cite{costmulti} is vital because the distinguishing features of various interference sources are most prominent in distinct domains: transient signatures like RADAR are distinct in the time domain, structured wideband signals like Wi-Fi are resolved in the frequency domain, and spectrally overlapping waveforms like LTE and 5G require \ac{csi} to separate. Integrating these features addresses the capacity limits of isolated models and significantly boosts classification accuracy \cite{IPFSCNN}, enabling a distributed, lightweight architecture to perform robustly within the 1 ms constraint.

\begin{figure*}
\centering
\includegraphics[width=\textwidth]{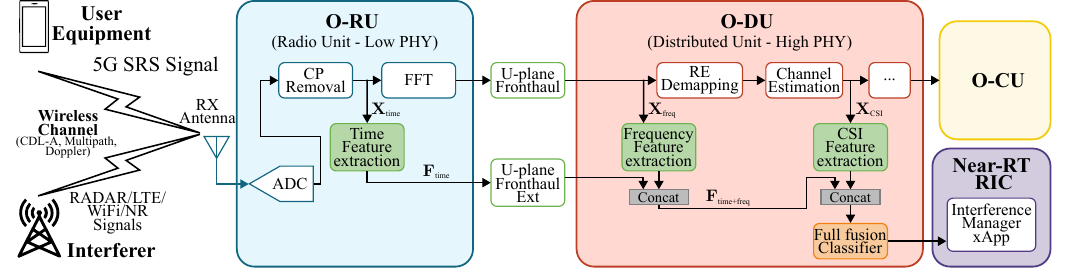}
\caption{System model of the proposed O-RAN-compliant uplink system, illustrating the extraction and processing of time, frequency, and CSI representations across the O-RU and O-DU.}
\label{fig:systemmodel}
\vspace{-0.4cm}
\end{figure*}
In this paper, we propose OSIRIS (O-RAN Split Inference for Radio Interference Sensing). To the best of our knowledge, this is the first lightweight model to align a multi-domain interference classifier directly with the \ac{oran} Split 7-2x \ac{phy} pipeline. Rather than operating on a single input, OSIRIS extracts time-domain samples, frequency-domain symbols, and \ac{csi} from the deterministic signal-processing chain. Dedicated extractors process these representations independently before fusing the latent features into a unified classification decision. The contributions of this paper are threefold:

\begin{itemize}
    \item We propose OSIRIS, a model that eliminates the computational burden of learning complex \ac{phy} transformations by utilizing the time, frequency, and \ac{csi} representations already computed within the \ac{oran} Split 7-2x pipeline.
    \item We introduce a staged pre-training strategy for multi-domain fusion. Optimizing domain-specific feature extractors before fusion improves training stability and convergence compared to standard random initialization. 
    \item We demonstrate that OSIRIS reduces inference latency by 9x and computational complexity by 11x against a monolithic baseline. This efficiency enables standard CPU deployment, satisfying the 1 ms 5G subframe constraint while maintaining high classification accuracy.
\end{itemize}

The remainder of this paper presents the system model, multi-domain model, dataset generation methodology, and an experimental evaluation demonstrating the efficiency of the proposed approach.

%% file: 2.system_architecture.tex
\input{acronyms}

\section{System Model \& Problem Formulation}

We consider an \ac{oran}-compliant uplink system operating under functional Split 7.2x, shown in \fref{fig:systemmodel}, comprising a single \ac{ue} transmitting a known 5G \ac{srs} to an \ac{ru}, potentially accompanied by an unknown interference signal and noise. Our objective is to detect and classify this interference (e.g., LTE, RADAR, Wi-Fi, \ac{nr}) and report the findings to the \ac{ric} via the E2 interface for an interference management xApp. The model input data is extracted across three representational domains directly from the standard \ac{phy} pipeline. First, the \ac{ru} performs \ac{adc} and \ac{cp} removal, yielding the time-domain sequence $\mathbf{X}_{\text{time}}$. Following the \ac{fft}, frequency-domain symbols are transmitted over the U-plane fronthaul to the \ac{du}, where they are immediately extracted as the frequency-domain sequence, $\mathbf{X}_{\text{freq}}$. Finally, after \ac{re} demapping and channel estimation at the \ac{du}, $\mathbf{X}_{\text{CSI}}$, the CSI-domain sequence is extracted. These three domains serve as input to our proposed OSIRIS model, fusing these complementary time, frequency, and \ac{csi} representations into a unified, low-complexity classifier. The following subsections detail the proposed OSIRIS architecture and its training methodology.

\subsection{Model Architecture}
 \begin{figure}
    \centering 
    \subfloat[Feature extraction backbone]{
    \includegraphics{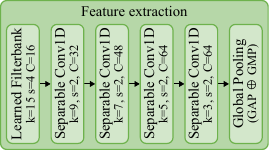}
    \label{fig:feature_arch}} 
    \subfloat[Classifier Head ]{
        \includegraphics{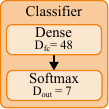}
        \label{fig:classifier_arch}}
    \caption{Detailed view of the proposed OSIRIS architecture, including (a) the domain-specific feature extraction backbone and (b) the classifier head.}
    \label{fig:architectures}
    \vspace{-0.3cm}
\end{figure}
Our proposed model employs a dedicated feature-extraction backbone for each representation domain, concatenating the resulting latent-space features before classifying the interference via a single classification head. The structure of the feature extraction backbone is detailed in \fref{fig:feature_arch}. We apply the same architectural setup across all domains: a learned filterbank with a kernel size of 15 and a stride of 4. This initial layer reduces the input sequence by a factor of 4, proportionally decreasing the computational load required in all subsequent layers compared to processing the raw, uncompressed signal. This filterbank is followed by four depthwise separable convolutional layers. Each of these layers performs a depthwise convolution on the input channels independently, followed by a pointwise convolution that combines the channel outputs.
In the \fref{fig:feature_arch}, $k$, $s$, and $C$ denote the kernel size, stride, and filter count, respectively. 
The extracted features $\mathbf{F}_{\text{time}}$, $\mathbf{F}_{\text{freq}}$, and $\mathbf{F}_{\text{CSI}}$ are concatenated and passed to the full fusion classifier shown in \fref{fig:classifier_arch} \cite{multimodal}, leveraging a late-fusion approach to effectively manage the heterogeneous domain features. This classifier consists of two fully connected layers: a hidden layer with 48 neurons and an output layer with 7 neurons, corresponding to the number of interference classes in the dataset. To balance model capacity and prevent over-fitting, both the hidden layer dimension and a dropout rate of 0.3 were selected via Bayesian optimization. A final Softmax layer outputs the posterior probabilities for the seven classes.

\subsection{Model training and selection}
To optimize the proposed model, we evaluate two distinct training strategies: OSIRIS$_{\text{rand}}$ and OSIRIS$_{\text{pre}}$. 
The first strategy, OSIRIS$_{\text{rand}}$, initializes the multi-domain model with random weights and trains the entire network end-to-end from scratch. While this approach is procedurally efficient because it requires only a single training pass, it forces the optimizer to learn both domain-specific feature extraction and cross-domain fusion simultaneously.
The second strategy, OSIRIS$_{\text{pre}}$, establishes robust individual feature representations prior to multi-domain integration. This process is executed in sequential stages. First, the time, frequency, and CSI feature extractors are pre-trained independently. During this phase, each domain-specific feature-extraction backbone is appended with an auxiliary classification head. The head consists of a hidden dense layer with 16 neurons followed by a 7-neuron output layer corresponding to the target interference classes. After training the domains separately, the best-performing feature extractors are selected, and their weights are transferred to the full OSIRIS model. To complete the training, we retrain the full model.

All models are implemented in TensorFlow and optimized using categorical cross-entropy loss. To prevent over-fitting and ensure optimal convergence, we apply a learning rate scheduler and an early stopping mechanism during each training phase. Furthermore, all architectural hyperparameters, including hidden layer dimensions, filter counts, and dropout rates, were determined via Bayesian optimization to maximize validation accuracy. To account for initialization variance, we train eight independent instances for every evaluated model and training stage. The instance that achieves the highest validation accuracy is subsequently selected for the final performance analysis.

\section{Dataset Generation}
To evaluate the proposed interference classification framework, a comprehensive synthetic dataset was developed to emulate a 5G-compliant spectral environment. The primary signal of interest is a 5G \ac{nr} \ac{srs}, synthesized in accordance with 3GPP TS 38.211 specifications. The signal uses a Zadoff-Chu sequence (root index $u=25$) mapped to a 100~MHz bandwidth with a 30~kHz subcarrier spacing, yielding an OFDM symbol with an \ac{fft} size of 4096. To ensure physical-layer realism, signals are propagated through a 3GPP TR 38.901 CDL-A multipath fading channel. This model incorporates six temporal taps with stochastic power and phase variations, complemented by a Doppler shift simulating mobility up to 30~m/s. 
The dataset encompasses seven interference classes generated via the MATLAB 5G Toolbox. Non-NR interferers include pulsed Radar signatures, LTE waveforms (\ac{RMC} R.7), and IEEE 802.11ax (Wi-Fi 6) \ac{HE} packets. The 5G \ac{nr} interferers are modeled across three configurations: 15~kHz numerology at 20~MHz ($\text{NR}_{\text{0,20}}$), and 30~kHz numerologies at 20~MHz ($\text{NR}_{\text{1,20}}$) and 40~MHz ($\text{NR}_{\text{1,40}}$). Additionally, a noise-only class is included to facilitate interference detection. These sources are aggregated with the \ac{srs} at \ac{sir} levels spanning $[-10, 10]$~dB, with \ac{snr} varied between $[-12, 20]$~dB.

For each simulation instance, three distinct domains are extracted: 4096 time-domain sequence samples, 4096 frequency-domain samples, and 1638 CSI-domain samples. The CSI domain samples are derived  by equalizing received subcarriers against the known Zadoff-Chu sequence.
The resulting dataset contains 1024 samples per SNR/SIR/Technology permutation, 9 SNR levels, 5 SIR levels, and 7 inference classes, resulting into total of 414,720 samples. The dataset is partitioned into training (75\%), validation (12.5\%), and hold-out test (12.5\%) sets for performance verification.

%% file: acronyms.tex
\begin{acronym}[HBCI]
\acro{6g}[6G]{6th Generation}
\acro{adc}[ADC]{Analog Digital Converter}
\acro{amc}[AMC]{Automatic Modulation Classification}
\acro{ap}[AP]{Access point}
\acro{awgn}[AWGN]{Additive White Gaussian Noise}
\acro{bn}[BN]{Batch Normalization}
\acro{bpsk}[BPSK]{Binary Phase Shift Keying}
\acro{cnn}[CNN]{Convolutional Neural Network}
\acro{cp}[CP]{Cyclic Prefix}
\acro{csi}[CSI]{Channel State Information}
\acro{cti}[CTI]{Cross-Technology Interference}
\acro{cu}[O-CU]{O-RAN Central Unit}
\acro{dApp}[dApp]{Distributed Apps}
\acro{dl}[DL]{Deep Learning}
\acro{dnn}[DNN]{Deep Neural Networks}
\acro{du}[O-DU]{O-RAN Distributed Unit}
\acro{ee}[EE]{Early Exiting}
\acro{egc}[EGC]{Equal Gain Combining}
\acro{er}[ER]{Early Rejection}
\acro{fb}[FB]{Feature-Based}
\acro{fft}[FFT]{Fast Fourier Transform}
\acro{flop}[FLOP]{Floating-Point Operation}
\acro{gap}[GAP]{Global Average Pooling}
\acro{gmp}{Global Max Pooling}
\acro{HE}[HE]{High Efficiency}
\acro{i}[I]{In-phase}
\acro{iq}[IQ]{In-phase/Quadrature}
\acro{lb}[LB]{Likelihood-Based}
\acro{mflop}[MFLOP]{Million FLOP}
\acro{ml}[ML]{Machine Learning}
\acro{mqam}[M-QAM]{M-ary Quadrature Amplitude Modulation}
\acro{nr}[NR]{New Radio}
\acro{oran}[O-RAN]{Open Radio Access Network}
\acro{phy}[PHY]{Physical layer}
\acro{q}[Q]{Quadrature}
\acro{qam}[QAM]{Quadrature Amplitude Modulation}
\acro{qpsk}[QPSK]{Quadrature Phase Shift Keying}
\acro{re}[RE]{Resource Elements}
\acro{relu}[ReLU]{Rectified Linear Unit}
\acro{rf}[RF]{Radio Frequency}
\acro{ric}[Near-RT RIC]{Near-Real-Time RAN Intelligent Controller}
\acro{RMC}[RMC]{Reference Measurement Channel}
\acro{rnn}[RNN]{Recurrent Neural Network}
\acro{ru}[O-RU]{O-RAN Radio Unit}
\acro{snr}[SNR]{Signal-to-Noise Ratio}
\acro{sir}[SIR]{Signal-to-Interference Ratio}
\acro{sota}[SotA]{State-of-the-Art}
\acro{srs}[SRS]{Sounding Reference Signal}
\acro{ue}[UE]{User Equipment}
\acro{wsn}[WSN]{Wireless Sensor Network}
\acro{wwee}[WWEE]{Width-Wise Early Exiting}
\end{acronym}

%% file: 3.Results.tex
\section{Results and Discussion}
We evaluate the training performance of the two proposed methods against a baseline ResNet that operates exclusively on the frequency-domain representation, which achieved the strongest performance among the evaluated single-domain models. ResNet was selected for its widespread use in wireless signal classification and strong performance in prior deep learning-based RF classification studies \cite{oshea}. While more lightweight models exist, ResNet provides a representative baseline for evaluating the efficiency–accuracy trade-off of the proposed approach. Our ResNet implementation consists of 8 residual stacks, 2 fully connected layers with 128 neurons each, and a final dense classification layer with 7 neurons. Consistent with the selection methodology for the OSIRIS model, we trained eight independent instances of the baseline and selected the best-performing model for our analysis.

\subsection{Training Performance}

The training performance of each strategy is evaluated using a batch size of 256 on an NVIDIA GeForce RTX 2080 Ti GPU. The evaluation metrics include the number of epochs, total training time, maximum classification accuracy, and training stability. Stability is defined as the difference ($\Delta$) between the maximum and minimum accuracy across the eight training iterations, where a lower $\Delta$ indicates greater robustness.
To assess the memory footprint, storage usage, and latency of the models, we converted them to the TensorFlow Lite format using its built-in tools. Benchmarking was conducted on a dual-socket Intel Xeon Gold 5220 system. We report the median and 95th percentile latency over 1,000 iterations, following a 50-iteration warm-up phase to account for CPU frequency scaling. 
The training efficiency and stability results are summarized in \tref{tab:model_performance}. For the OSIRIS$_{\text{pre}}$ strategy, the total training time of 142 minutes includes pre-training for the time (26 min), frequency (26 min), and CSI (19 min) sub-models, followed by the multi-domain fusion stage (71 min).
\begin{table}
\centering
\caption{Training Performance and Cost}
\label{tab:model_performance}
\small
\setlength{\tabcolsep}{4pt} 
\begin{tabular}{lcccccc}
\toprule
\textbf{Model} & \textbf{Epochs} & \textbf{Time}\textbf{(min)} & \textbf{Max Acc.}\textbf{(\%)} & \textbf{Stability}\textbf{(\%)} \\ 
\midrule
OSIRIS$_{\text{pre}}$        & \textbf{57} & 142  & 80.22 & \textbf{2.14} \\
OSIRIS$_{\text{rand}}$      & 64 & \textbf{76}  & 78.72 & 6.43 \\
ResNet          & 64 & 187  & \textbf{84.86 }& 3.18 \\
\bottomrule
\end{tabular}
\vspace{-0.4cm}
\end{table}
Although the staged pre-training requires a higher initial time investment than the randomly initialized OSIRIS${\text{rand}}$ method (76 min), it yields a significant improvement in stability and convergence speed. The OSIRIS${\text{pre}}$ model achieves a stability variance of 2.14\% and converges in 57 epochs, compared to 6.43\% and 64 epochs for the OSIRIS$_{\text{rand}}$ approach. This shows that independent pre-training provides a stable foundation for the fusion layers. By allowing the network to first learn the distinct features of each domain, the final training stage can focus entirely on optimizing cross-domain synergy, avoiding the unpredictable performance swings of training from scratch. The ResNet requires the longest training time (187 min) and 64 epochs to converge, despite being a single-stage training process. This extended duration directly reflects its higher computational complexity and larger parameter space. While it achieves the highest overall accuracy (84.86\%), its stability (3.18\%) falls between the pre-trained and randomly initialized OSIRIS models.

\begin{table}
    \centering
    \caption{Inference Performance and Cost}
    \label{tab:single_modality_results}
    \setlength{\tabcolsep}{5pt} 
    \begin{tabular}{l cc cc}
        \toprule
         \multirow{2}{*}{\textbf{Model}} & \multicolumn{2}{c}{\textbf{Latency} ($\mu s$)} & \multicolumn{2}{c}{\textbf{Complexity}} \\
        \cmidrule(lr){2-3} \cmidrule(lr){4-5}
         & Median & $95^{th}$ \% & MFLOPs & kParams \\
        \midrule
        
          ResNet  & 3761 & 4266 & 55.14 & 115.6 \\
          OSIRIS  & \textbf{398} & \textbf{494} & \textbf{4.9} & \textbf{52.7} \\
        \bottomrule
    \end{tabular}
    \vspace{-0.4cm}
\end{table}

\tref{tab:single_modality_results} highlights the operational trade-off of the proposed model. While ResNet achieves a slightly higher peak accuracy, it requires 55.14 \acp{mflop} and 115.6 kParams, incurring a median latency of 3761 $\mu s$ and a 95th percentile latency of 4266 $\mu s$. In contrast, the OSIRIS model reduces this computational overhead. It requires only 4.9 MFLOPs and 52.7 kParams, achieving a median latency of 398 $\mu s$ and maintaining a 95th percentile latency of 494 $\mu s$. Analyzing this execution overhead within the context of the O-RAN functional split, approximately 42\% of both the processing latency and computational load occurs at the \ac{ru} for the time-domain sequence. The remaining 58\% is handled at the \ac{du}, divided between the frequency (42\%) and CSI (16\%) representations. Ultimately, this reliable sub-millisecond execution confirms that the proposed model successfully balances robust classification with the strict latency constraints required for real-time deployment. This behavior underscores the core compromise of the proposed method: explicitly mapping multi-domain fusion to the O-RAN split reduces computational latency, the shallower feature-extraction layers degrade discriminative resolution for spectrally identical bandwidths.

\begin{figure*}
    \begin{minipage}{0.33\textwidth}
        \centering
        \vspace{1cm} 
        \input{figures/tex/Graph_sir} 
        \caption{Classification accuracy across varying SIR levels for the ResNet and the proposed OSIRIS variants.}
        \label{fig:sir_performance}
    \end{minipage}
    \hfill
    \begin{minipage}{0.65\textwidth}
        \centering
        \subfloat[ResNet]{%
            \input{figures/tex/cm_resnet}%
        }
        \hfill
        \subfloat[$\text{OSIRIS}_{\text{pre}}$]{%
            \input{figures/tex/cm_pretrained}%
        }
        \hfill
        \subfloat[OSIRIS$_\text{rand}$]{%
            \input{figures/tex/cm_rand}%
        }
        \caption{Confusion matrices detailing the classification performance of (a) the single-domain ResNet, (b) the pre-trained OSIRIS model, and (c) the randomly initialized OSIRIS model.}
        \label{fig:all_cms}
    \end{minipage}
  \vspace{-0.49cm}
\end{figure*}

\fref{fig:sir_performance} provides a detailed comparison of model robustness under varying \ac{sir} conditions. As expected, classification accuracy is highest at lower SIR levels (-10 dB), where the relative interference energy is dominant and signatures are more pronounced. We observe a characteristic performance decay across all models as the SIR increases from -10 dB to 10 dB. Notably, the OSIRIS$_{\text{pre}}$ variant consistently outperforms the OSIRIS$_{\text{rand}}$ by a margin of 1.4\% to 1.7\% across the entire evaluation range.

To further analyze the classification behavior, we present the confusion matrices for ResNet, OSIRIS$_{\text{pre}}$, and OSIRIS$_{\text{rand}}$ in \fref{fig:all_cms}. The matrices reveal a significant confusion cluster between LTE and $\text{NR}_{\text{1,20}}$ signals across all models, likely due to overlapping spectral characteristics. However, a comparative analysis shows that OSIRIS$_{\text{pre}}$ demonstrates superior discriminative power compared to the OSIRIS$_{\text{rand}}$. Specifically, the confusion between $\text{NR}_{\text{1,20}}$ and LTE is reduced from 20.6\% to 16.9\% using pre-trained domain modules. This qualitative improvement confirms that staged pretraining specifically strengthens the model's ability to disambiguate highly similar interference waveforms. Furthermore, the matrices highlight the specific operational trade-offs of the lightweight OSIRIS model. All models demonstrate near-optimal classification for distinct signatures such as Noise (exceeding 97.9\%) and RADAR (exceeding 96.4\%). Similarly, the wider 40 MHz bandwidth of $\text{NR}_{\text{1,40}}$ provides sufficient spectral distinction for robust detection across all configurations (86.5\% to 89.7\%). The primary limitation of the OSIRIS model emerges within the 20 MHz waveforms. While the monolithic ResNet maintains a distinct advantage in classifying LTE (78.1\% accuracy), OSIRIS$_{\text{pre}}$ achieves 63.9\%, exhibiting a higher tendency to misclassify LTE as $\text{NR}_{\text{1,20}}$ (15.5\% versus ResNet's 10.1\%).

\subsection{Cross-domain Collaboration}
\begin{figure*}
    \centering
    \subfloat[RADAR]{%
        \label{fig:det_radar}%
        \input{figures/tex/Graph_detailed_radar}%
    }
    \hfill
    \subfloat[Wi-Fi]{%
        \label{fig:det_wifi}%
        \input{figures/tex/Graph_detailed_wifi}%
    }
    \hfill
    \subfloat[$\text{NR}_{\text{1,20}}$]{%
        \label{fig:det_nr}%
        \input{figures/tex/Graph_detailed_nr}%
    }
    \caption{Classification accuracy versus SIR for (a) RADAR, (b) Wi-Fi, and (c) $\text{NR}_{\text{1,20}}$ across isolated domain models and the OSIRIS model.}
    \label{fig:crossdomain_detailed}
    \vspace{-0.4cm}
\end{figure*}
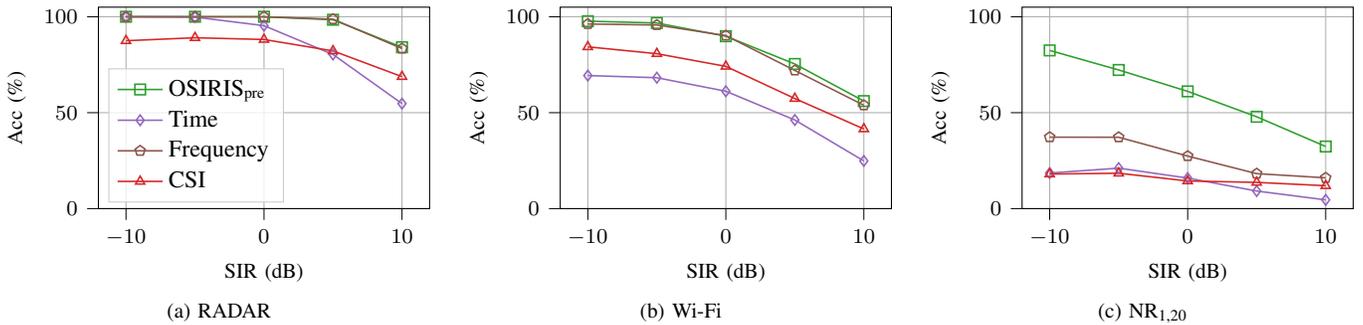

To evaluate cross-domain collaboration, we analyzed classification accuracy across varying SIR levels, shown in \fref{fig:crossdomain_detailed}, for three distinct interference classes: RADAR, Wi-Fi, and $\text{NR}_{\text{1,20}}$. The data demonstrates that the OSIRIS model automatically relies on the most informative domain for each signal type. 
For distinct transient signals such as RADAR, the multi-domain fusion achieves 100\% accuracy at -10 dB SIR, matching the optimal performance of the time and frequency domains. For structured wideband signals such as Wi-Fi, the frequency domain provides the strongest individual features (96.2\% at -10 dB). The proposed OSIRIS actively tracks this, reaching 97.7\% accuracy.
This result highlights the primary advantage of multi-domain fusion. 
When evaluated individually, none of the representation domains provides sufficient discriminative power for NR1,20 interference. However, by combining complementary features across domains, the multi-domain OSIRIS model resolves the classification task, achieving 82.4\% accuracy. This demonstrates that multi-domain fusion can exploit weak but complementary signals that remain inaccessible to lightweight single-domain classifiers.

%% file: figures/tex/Graph_sir.tex
\begin{tikzpicture}

\definecolor{crimson2143940}{RGB}{214,39,40}
\definecolor{darkgray176}{RGB}{176,176,176}
\definecolor{darkorange25512714}{RGB}{255,127,14}
\definecolor{forestgreen4416044}{RGB}{44,160,44}
\definecolor{goldenrod18818934}{RGB}{188,189,34}
\definecolor{gray127}{RGB}{127,127,127}
\definecolor{lightgray204}{RGB}{204,204,204}
\definecolor{mediumpurple148103189}{RGB}{148,103,189}
\definecolor{orchid227119194}{RGB}{227,119,194}
\definecolor{sienna1408675}{RGB}{140,86,75}
\definecolor{steelblue31119180}{RGB}{31,119,180}

\begin{axis}[
legend cell align={left},
legend style={
  fill opacity=0.8,
  draw opacity=1,
  text opacity=1,
  at={(0.03,0.03)},
  anchor=south west,
  draw=lightgray204
},
tick align=outside,
tick pos=left,
x grid style={darkgray176},
xlabel={SIR (dB)},
xmajorgrids,
xmin=-12, xmax=12,
xtick style={color=black},
y grid style={darkgray176},
ylabel={Accuracy (\%)},
ymajorgrids,
ymin=45, ymax=105,
ytick style={color=black},
width=\linewidth,
height=0.75\linewidth,
]

\addplot [semithick, mark=o,mark repeat=1,mark phase = 1,darkorange25512714]
table [x=SIR, y=ResNetavg, col sep=comma] {figures/csv/SIR.csv};
\addlegendentry{ResNet}

\addplot [semithick, forestgreen4416044,mark=square,mark repeat=1, mark phase=1]
table [x=SIR, y=Multimodalavg, col sep=comma] {figures/csv/SIR.csv};
\addlegendentry{OSIRIS$_\text{pre}$}

\addplot [semithick,mark=star,mark repeat=1,mark phase = 1, steelblue31119180] 
table [x=SIR, y=oneshotavg, col sep=comma] {figures/csv/SIR.csv};
\addlegendentry{OSIRIS$_\text{rand}$}







\end{axis}
\end{tikzpicture}

%% file: figures/tex/cm_resnet.tex
\begin{tikzpicture}

\definecolor{darkgray176}{RGB}{176,176,176}
\definecolor{darkslategray38}{RGB}{38,38,38}

\begin{axis}[
tick align=outside,
tick pos=left,
x grid style={darkgray176},
xlabel={Predicted Label},
xmin=-0.5, xmax=6.5,
xtick style={color=black},
xtick={0,1,2,3,4,5,6},
xticklabel style={rotate=45.0},
xticklabels={Noise,LTE,RADAR,Wi-Fi,$\text{NR}_{\text{0,20}}$,$\text{NR}_{\text{1,20}}$,$\text{NR}_{\text{1,40}}$},
xticklabel style={font=\scriptsize,overlay,},
y dir=reverse,
y grid style={darkgray176},
ylabel={True Label},
ymin=-0.5, ymax=6.5,
ytick style={color=black},
ytick={0,1,2,3,4,5,6},
yticklabels={Noise,LTE,RADAR,Wi-Fi,$\text{NR}_{\text{0,20}}$,$\text{NR}_{\text{1,20}}$,$\text{NR}_{\text{1,40}}$},
yticklabel style={font=\scriptsize,},
yticklabel style={rotate=45.0},
scale only axis,
width=0.22\linewidth,
height=0.22\linewidth,
]

\addplot graphics [includegraphics cmd=\pgfimage,xmin=-0.5, xmax=6.5, ymin=6.5, ymax=-0.5] {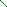};
\addplot[
    only marks,     
    nodes near coords, 
    point meta=explicit symbolic, 
    visualization depends on={value \thisrow{textcolor} \as \mycolor},
    nodes near coords style={
        scale=0.75,      
        anchor=center,  
        rotate=0.0,     
        color=\mycolor   
    }
] table [x=x, y=y, meta=val, col sep=comma] {figures/csv/cm_ResNet.csv};
\end{axis}

\end{tikzpicture}

%% file: figures/tex/cm_pretrained.tex
\begin{tikzpicture}

\definecolor{darkgray176}{RGB}{176,176,176}
\definecolor{darkslategray38}{RGB}{38,38,38}

\begin{axis}[
tick align=outside,
tick pos=left,
x grid style={darkgray176},
xlabel={Predicted Label},
xmin=-0.5, xmax=6.5,
xtick style={color=black},
xtick={0,1,2,3,4,5,6},
xticklabel style={rotate=45.0},
xticklabels={Noise,LTE,RADAR,Wi-Fi,$\text{NR}_{\text{0,20}}$,$\text{NR}_{\text{1,20}}$,$\text{NR}_{\text{1,40}}$},
xticklabel style={font=\scriptsize,overlay,},
y dir=reverse,
y grid style={darkgray176},
ylabel={},
ymin=-0.5, ymax=6.5,
ytick style={color=black},
ytick={0,1,2,3,4,5,6},
yticklabels={\empty},
yticklabel style={font=\scriptsize,},
yticklabel style={rotate=45.0},
scale only axis,
width=0.22\linewidth,
height=0.22\linewidth,
trim axis left,        
trim axis right,  
]

\addplot graphics [includegraphics cmd=\pgfimage,xmin=-0.5, xmax=6.5, ymin=6.5, ymax=-0.5] {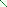};
\addplot[
    only marks,     
    nodes near coords, 
    point meta=explicit symbolic, 
    visualization depends on={value \thisrow{textcolor} \as \mycolor},
    nodes near coords style={
        scale=0.75,      
        anchor=center,  
        rotate=0.0,     
        color=\mycolor   
    }
] table [x=x, y=y, meta=val, col sep=comma] {figures/csv/cm_pretrained.csv};
\end{axis}

\end{tikzpicture}

%% file: figures/tex/cm_rand.tex
\begin{tikzpicture}

\definecolor{darkgray176}{RGB}{176,176,176}
\definecolor{darkslategray38}{RGB}{38,38,38}

\begin{axis}[
tick align=outside,
tick pos=left,
x grid style={darkgray176},
xlabel={Predicted Label},
xmin=-0.5, xmax=6.5,
xtick style={color=black},
xtick={0,1,2,3,4,5,6},
xticklabel style={rotate=45.0},
xticklabels={Noise,LTE,RADAR,Wi-Fi,$\text{NR}_{\text{0,20}}$,$\text{NR}_{\text{1,20}}$,$\text{NR}_{\text{1,40}}$},
xticklabel style={font=\scriptsize,overlay,},
y dir=reverse,
y grid style={darkgray176},
ylabel={},
ymin=-0.5, ymax=6.5,
ytick style={color=black},
ytick={0,1,2,3,4,5,6},
yticklabels={\empty},
yticklabel style={font=\scriptsize,},
yticklabel style={rotate=45.0},
scale only axis,
width=0.22\linewidth,
height=0.22\linewidth,
trim axis left,        
trim axis right,  
]

\addplot graphics [includegraphics cmd=\pgfimage,xmin=-0.5, xmax=6.5, ymin=6.5, ymax=-0.5] {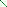};
\addplot[
    only marks,     
    nodes near coords, 
    point meta=explicit symbolic, 
    visualization depends on={value \thisrow{textcolor} \as \mycolor},
    nodes near coords style={
        scale=0.75,      
        anchor=center,  
        rotate=0.0,     
        color=\mycolor   
    }
] table [x=x, y=y, meta=val, col sep=comma] {figures/csv/cm_random.csv};
\end{axis}

\end{tikzpicture}

%% file: figures/tex/Graph_detailed_radar.tex
\begin{tikzpicture}

\definecolor{crimson2143940}{RGB}{214,39,40}
\definecolor{darkgray176}{RGB}{176,176,176}
\definecolor{darkorange25512714}{RGB}{255,127,14}
\definecolor{forestgreen4416044}{RGB}{44,160,44}
\definecolor{goldenrod18818934}{RGB}{188,189,34}
\definecolor{gray127}{RGB}{127,127,127}
\definecolor{lightgray204}{RGB}{204,204,204}
\definecolor{mediumpurple148103189}{RGB}{148,103,189}
\definecolor{orchid227119194}{RGB}{227,119,194}
\definecolor{sienna1408675}{RGB}{140,86,75}
\definecolor{steelblue31119180}{RGB}{31,119,180}

\begin{axis}[
legend cell align={left},
legend columns=1,
legend style={
  fill opacity=0.8,
  draw opacity=1,
  text opacity=1,
  at={(0.03,0.03)},
  anchor=south west,
  draw=lightgray204,
},
tick align=outside,
tick pos=left,
x grid style={darkgray176},
xlabel={SIR (dB)},
xmajorgrids,
xmin=-12, xmax=12,
xtick style={color=black},
y grid style={darkgray176},
ylabel={Acc (\%)},
ymajorgrids,
ymin=0, ymax=105,
ytick style={color=black},
width=0.33\linewidth,
height=0.235\linewidth,
]


\addplot [semithick, mark=square,mark repeat=1,mark phase = 1,forestgreen4416044]
table [x=SIR, y=MultimodalRADAR, col sep=comma] {figures/csv/Graph_SingleModal_SIR_detailed.csv};
\addlegendentry{$\text{OSIRIS}_\text{pre}$ }

\addplot [semithick, mediumpurple148103189,mark=diamond,mark repeat=1, mark phase=1]
table [x=SIR, y=timeRADAR, col sep=comma] {figures/csv/Graph_SingleModal_SIR_detailed.csv};
\addlegendentry{$\text{Time}_\text{}$}

\addplot [semithick, sienna1408675,mark=pentagon,mark repeat=1, mark phase=1]
table [x=SIR, y=freqRADAR, col sep=comma] {figures/csv/Graph_SingleModal_SIR_detailed.csv};
\addlegendentry{$\text{Frequency}_\text{}$}

\addplot [semithick, mark=triangle,mark repeat=1,mark phase = 1,crimson2143940]
table [x=SIR, y=csiRADAR, col sep=comma] {figures/csv/Graph_SingleModal_SIR_detailed.csv};
\addlegendentry{$\text{CSI}_\text{}$ }






\end{axis}
\end{tikzpicture}

%% file: figures/tex/Graph_detailed_wifi.tex
\begin{tikzpicture}

\definecolor{crimson2143940}{RGB}{214,39,40}
\definecolor{darkgray176}{RGB}{176,176,176}
\definecolor{darkorange25512714}{RGB}{255,127,14}
\definecolor{forestgreen4416044}{RGB}{44,160,44}
\definecolor{goldenrod18818934}{RGB}{188,189,34}
\definecolor{gray127}{RGB}{127,127,127}
\definecolor{lightgray204}{RGB}{204,204,204}
\definecolor{mediumpurple148103189}{RGB}{148,103,189}
\definecolor{orchid227119194}{RGB}{227,119,194}
\definecolor{sienna1408675}{RGB}{140,86,75}
\definecolor{steelblue31119180}{RGB}{31,119,180}

\begin{axis}[
legend cell align={left},
legend columns=4,
legend style={
  fill opacity=0.8,
  draw opacity=1,
  text opacity=1,
  at={(0.03,0.03)},
  anchor=south west,
  draw=lightgray204,
  overlay
},
tick align=outside,
tick pos=left,
x grid style={darkgray176},
xlabel={SIR (dB)},
xmajorgrids,
xmin=-12, xmax=12,
xtick style={color=black},
y grid style={darkgray176},
ylabel={Acc (\%)},
ymajorgrids,
ymin=0, ymax=105,
ytick style={color=black},
width=0.33\linewidth,
height=0.235\linewidth,
]

\addplot [semithick, mark=square,mark repeat=1,mark phase = 1,forestgreen4416044]
table [x=SIR, y=MultimodalWIFI, col sep=comma] {figures/csv/Graph_SingleModal_SIR_detailed.csv};

\addplot [semithick, mediumpurple148103189,mark=diamond,mark repeat=1, mark phase=1]
table [x=SIR, y=timeWIFI, col sep=comma] {figures/csv/Graph_SingleModal_SIR_detailed.csv};

\addplot [semithick, sienna1408675,mark=pentagon,mark repeat=1, mark phase=1]
table [x=SIR, y=freqWIFI, col sep=comma] {figures/csv/Graph_SingleModal_SIR_detailed.csv};

\addplot [semithick, mark=triangle,mark repeat=1,mark phase = 1,crimson2143940]
table [x=SIR, y=csiWIFI, col sep=comma] {figures/csv/Graph_SingleModal_SIR_detailed.csv};







\end{axis}
\end{tikzpicture}

%% file: figures/tex/Graph_detailed_nr.tex
\begin{tikzpicture}

\definecolor{crimson2143940}{RGB}{214,39,40}
\definecolor{darkgray176}{RGB}{176,176,176}
\definecolor{darkorange25512714}{RGB}{255,127,14}
\definecolor{forestgreen4416044}{RGB}{44,160,44}
\definecolor{goldenrod18818934}{RGB}{188,189,34}
\definecolor{gray127}{RGB}{127,127,127}
\definecolor{lightgray204}{RGB}{204,204,204}
\definecolor{mediumpurple148103189}{RGB}{148,103,189}
\definecolor{orchid227119194}{RGB}{227,119,194}
\definecolor{sienna1408675}{RGB}{140,86,75}
\definecolor{steelblue31119180}{RGB}{31,119,180}

\begin{axis}[
legend cell align={left},
legend columns=4,
legend style={
  fill opacity=0.8,
  draw opacity=1,
  text opacity=1,
  at={(0.03,0.03)},
  anchor=south west,
  draw=lightgray204,
},
tick align=outside,
tick pos=left,
x grid style={darkgray176},
xlabel={SIR (dB)},
xmajorgrids,
xmin=-12, xmax=12,
xtick style={color=black},
y grid style={darkgray176},
ylabel={Acc (\%)},
ymajorgrids,
ymin=0, ymax=105,
ytick style={color=black},
width=0.33\linewidth,
height=0.235\linewidth,
]

\addplot [semithick, mark=square,mark repeat=1,mark phase = 1,forestgreen4416044]
table [x=SIR, y=MultimodalNRmu120, col sep=comma] {figures/csv/Graph_SingleModal_SIR_detailed.csv};

\addplot [semithick, mediumpurple148103189,mark=diamond,mark repeat=1, mark phase=1]
table [x=SIR, y=timeNRmu120, col sep=comma] {figures/csv/Graph_SingleModal_SIR_detailed.csv};

\addplot [semithick, sienna1408675,mark=pentagon,mark repeat=1, mark phase=1]
table [x=SIR, y=freqNRmu120, col sep=comma] {figures/csv/Graph_SingleModal_SIR_detailed.csv};

\addplot [semithick, mark=triangle,mark repeat=1,mark phase = 1,crimson2143940]
table [x=SIR, y=csiNRmu120, col sep=comma] {figures/csv/Graph_SingleModal_SIR_detailed.csv};








\end{axis}
\end{tikzpicture}

%% file: 4.Conclusion.tex
\section{Conclusion}
In this paper, we propose OSIRIS, a lightweight multi-domain interference classification architecture designed for the O-RAN \ac{phy}. Our evaluations demonstrate that OSIRIS reduces computational complexity by 11x and inference latency by 9x compared to a monolithic ResNet baseline, achieving a median latency of approximately 398 $\mu s$ by fusing knowledge of multiple domains. Furthermore, we showed that a staged pre-training strategy, OSIRIS$_{\text{pre}}$, enhances training stability compared to a random initialisation OSIRIS$_{\text{rand}}$, and improves the differentiation of spectrally overlapping waveforms, such as LTE and $\text{NR}_{\text{1,20}}$, compared to single-domain models. Ultimately, this work illustrates that explicitly aligning neural network architectures with complementary \ac{phy} domain representations enables the low-latency, computation efficient interference intelligence required for real-time \ac{dApp} deployments in next-generation networks.